# Vortices associated with the wave function of a single electron emitted in slow ion-atom collisions

L. Ph. H. Schmidt[1], C. Goihl[1], D. Metz[1], H. Schmidt-Böcking[1], R. Dörner[1], S. Yu. Ovchinnikov[2,3,4], J. H. Macek[4], and D. R. Schultz[2]

[1]*Institut für Kernphysik, Goethe-Universität, 60438 Frankfurt am Main, Germany*
[2]*Department of Physics, University of North Texas, Denton, Texas 76203, USA*
[3]*Ioffe Physical Technical Institute, St. Petersburg, 194021 Russia*
[4]*Department of Physics and Astronomy, University of Tennessee, Knoxville, Tennessee 37496, USA*

We present measurements and calculations of the momentum distribution of electrons emitted during the ion-atom collision 10 keV/u $He^{2+}$ + He ➔ $He^+$ + $He^{2+}$ + $e^-$, which show rich structures for ion scattering angles above 2 mrad arising dominantly from two-electron states. Our calculations reveal that minima in the measured distributions are zeroes in the electronic probability density resulting from vortices in the electronic current.

Molecular bonds are formed and broken in slow ion-atom collisions on a sub-femtosecond time scale. Upon bond breaking, the electrons forming the bond can be promoted to the continuum where the square of their wave function can be measured. A significant and unexpected electron dynamics in this process was predicted in the 1980's [1] to strand electrons in the saddle point region of the two-center potential between the receding target and projectile ions. Using the reaction microscopy technique, it was experimentally [2,3] confirmed that this process, for example, cools the ejected electrons so that they possess much narrower transverse momentum distributions than those of the initial atomic states they emerged from. Furthermore, the electron emission pattern showed unexpected structures that can be assigned to the electron dynamics on the saddle [3,4,5] or are predicted to arise from quantized transfer of angular momentum from the nuclei to the electron via formation of vortices in the electronic probability current. These predicted vortices in the current correspond to zeros of the electronic probability density [6,7]. A key discovery of the reaction microscope experiments and the corresponding theoretical descriptions has been that electrons ejected in atomic processes image high-lying transient quasi-molecular states. In this Letter we use this transient state imaging to demonstrate that the predicted zeros are associated with the vortices.

Experimentally the vortices should show up as small regions of minimal density in the final state electron momentum distribution if the impact parameter could be inferred experimentally with sufficient precision by measuring the scattering angle. Resolving these minima, which have small extension in momentum and very small (ideally zero) magnitude, was impossible with the resolution and statistical significance of all previous experiments. Therefore, to demonstrate these vortices experimentally, we completely redesigned the Frankfurt reaction microscope setup that was earlier used to measure electron emission in slow ion-atom collisions [4]. We measured the transfer ionization in the process 10 keV/u $He^{2+}$ + He ➔ $He^+$ + $He^{2+}$ + $e^-$ (see experimental geometry illustrated in Figure 1). To do so, a new cold target recoil ion spectrometer (COLTRIMS) with a 124 mm diameter position-sensitive microchannel plate detector in the electron arm of the spectrometer was constructed. No magnetic field is used and the electron detector is shifted forward [8] in the direction of the outgoing projectile to collect electrons with velocities between the target and the projectile ion over the whole active surface. The extraordinary electron momentum resolution required for this study requires a low electric field of only 0.09 V/mm. We reach an electron momentum resolution between 0.01 and 0.02 a.u. FWHM depending on the direction of emission (a.u. = atomic units, defined by $m_e = \hbar = e = 4\pi\varepsilon_0 = 1$). Extracting recoiling ions with momenta up to 10 a.u. from the reaction region with such a low field spreads them in the ion arm of our spectrometer over more than 150 mm. By using additional electrostatic lenses we refocus them onto a detector with an 80 mm active surface. This yields a recoil ion momentum resolution of 0.1 a.u. FWHM.

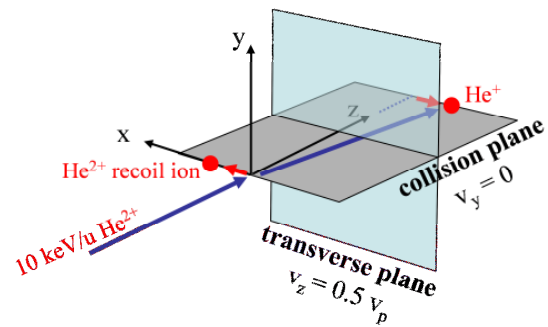

Figure 1. Coordinate system defined by the nuclear motion that is used to present the emitted electron distribution of the transfer ionization reaction 10 keV/u $He^{2+}$ + He ➔ $He^+$ + $He^{2+}$ + $e^-$. The collision plane $(x,z)$ is defined by the nuclear motion with the $z$ axis being the direction of impact. The real projectile deflection angles are much smaller than sketched here. The $y$ axis is normal to the collision plane. The electron emission pattern is found to be symmetric with respect to the collision plane $(y = 0)$ and therefore the experimental data have been mirrored to decrease the statistical errors.

To unambiguously associate the detected regions of minimal density in the ejected electron spectrum with vortices that survive from the near-collision to asymptotic distances, and to reveal the mechanisms responsible for the observed behavior of the measured momentum distribution, theoretical interpretation is also necessary. Since a full solution of the time-dependent Schrödinger equation (TDSE) for two fully correlated electrons propagated to infinity in a six-dimensional space is still a great challenge, particularly for a range of impact energies and geometries required to explore the conditions covered by the experiment, an interpretative rather than fully *ab initio* approach is needed. Our previous fully correlated, but four-dimensional ("flatland"), solution of the TDSE [7], which predicted the presence of vortices for transfer ionization, treated only the longitudinal (collision) plane for a single impact parameter. The present work has shown that the non-sharp relationship of impact parameter and momentum transfer [9] blurs out the vortices in this plane and that the transverse plane, not represented in the flatland model, displays the presence of the vortices.

From a theoretical point of view, during a slow ion-atom collision the transient states largely have the form of the quasi-molecular orbitals that adapt adiabatically to the field of the moving nuclei. The nuclear dynamics then induces transitions between these orbitals causing the top-of-barrier promotion to highly excited quasi-molecular states. In particular, the probability of electron emission to the continuum can be described by the hidden crossing (HC) approach [10], which identifies the pathway to the continuum via a sequence of branch points of the quasi-molecular potential energy surface extended to complex values of the internuclear distances. This picture led to identification of the main features of the observed [2] electron emission pattern in a one-electron transition as a coherent superposition of quasi-molecular states with σ and π symmetry [5]. Description of transfer ionization requires an analogous two-electron HC (2eHC) model [11] to identify the quasi-molecular states dominantly contributing to the electron promotion, and to describe the population of $\delta$ states, which affect the electron emission pattern in this case. These two-electron transitions occur at relatively small internuclear separations of about $R_\delta = 1.1$ a.u.

As we have noted, a full solution of the TDSE for two active electrons is a not feasible, so we therefore use a hybrid model in which we treat the full two-electron dynamics via the 2eHC approach when the two nuclei are close (within 10 a.u.) and then propagate to large distances the one-electron wave function that describes the escape to the continuum using the regularized and scaled lattice TDSE method (RLTDSE) [12,13,14]. Specifically, in the inner region the impact parameter-dependent two-electron wave function is written as a superposition of Sturmian basis functions

$$\psi_i = \sum_{fj}{}' A_{fj}^{(i)}(b)\, \phi_f(r_1)\, \phi_j^{(i)}(r_2)$$

where the prime on the sum indicates that the product function $\phi_f(r_1)\phi_j^{(i)}(r_2)$ is chosen to correspond to physical initial conditions where both electrons are localized on the He target. The Sturmian products correspond to three different initial molecular orbital channels of the quasi-molecule $\mathrm{He}_2^{2+}$: $(1s\sigma_g)^2$, $(2p\sigma_u)^2$, and $(1s\sigma_g, 2p\sigma_u)$. Each channel, for a given impact parameter $b$, is propagated to 10 a.u. using 2eHC where paths (*i*) are in the complex plane of internuclear separation yielding a wave function of the form given above.

A two-dimensional Fourier transformation on $A_{fj}^{(i)}(b)$ then gives an amplitude $A_{fj}^{(i)}(p_\perp)$, and hence a wave function that is a function of $p_\perp$. For a given channel one can compute the potential $V_{eff}(r_2) = \langle f | V | f \rangle$ that the second electron moves in for a given final channel *f* corresponding to an electron in a molecular orbital ground state. The impact parameter used in this step $b = b_{fj}$ is the stationary phase point of the integrand in the Fourier integral. The one-electron effective potential is then used to propagate $\phi_j^{(i)}(r_2)$ to asymptotic distances (~$10^5$ a.u.) using the RLTDSE method [13] for a straight-line trajectory, where, because the imaging theorem [15] relates the expanding coordinate space to the measured electron momenta, we can directly compare the results of our RLTDSE approach to the experimental data. That is, the absolute value squared of the sum

$$\sum_j \left( \sum_{if} A_{fj}^{(i)}(p_\perp)\, u_j^{(i)}(k) \right),$$

where $u_j^{(i)}(k)$ is the ionization amplitude obtained from the basis functions $\phi_j^{(i)}(r_2)$, gives the computed ejected electron distributions for this hybrid approach.

While the relative phases of the amplitudes $A_{fj}^{(i)}(p_\perp)$ are known in principle from this result, in practice the phases in the 2eHC model are approximate. Therefore, we slightly correct the phases between the three initial channels to better fit the measured data. We used twelve final channel functions to describe the longitudinal distribution because only these twelve are populated with significant probability at 10 a.u. For the transverse distribution only two of these twelve states were dominant and needed due to the symmetry of the states in that plane. The two-state model also had the benefit of allowing us to identify in the simplest way the mechanisms responsible for the features of the measured spectrum of electrons (e.g., zeros associated with vortices). In this case only one slowly varying phase was required to be fit.

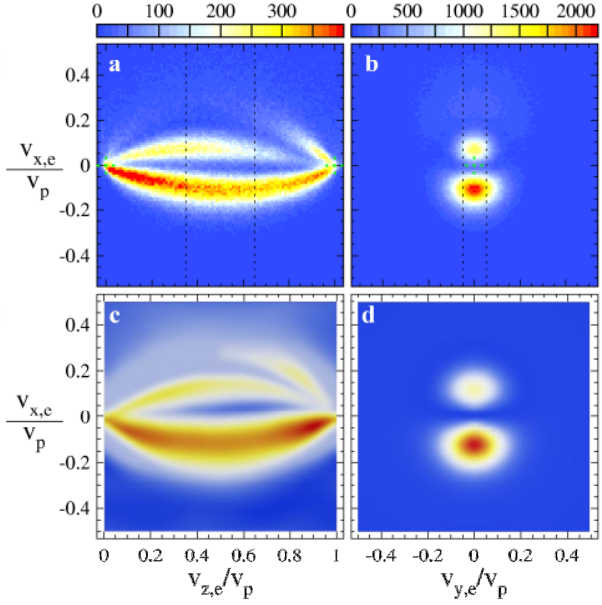

Figure 2. Measured (a and b) and predicted (c and d) ejected election velocity distribution for the transfer ionization process 10 kV/u He$^{2+}$ + He ➔ He$^+$ + He$^{2+}$ + e$^-$ at an impact velocity $v_p$= 0.63 a.u. and distant collisions experimentally selected by scattering angles below 1.25 mrad. The experimental distribution in the collision plane (a) contains events with out-of-plane velocities $v_{y,e}$ up to 0.05 $v_p$. The distribution in the (x,y) plane perpendicular to the direction of impact (b) is integrated from $v_{z,e}/v_p$ = 0.35 to 0.65. (c) 2eHC-RLTDSE calculation for an impact parameter $b$ = 1.4 a.u. and $v_{z,e}/v_p$ = 0.5 considering only the two dominant states.

The momentum distribution of the emitted electron in the collision plane (the longitudinal distribution), which is defined by the direction of impact and the nuclear momentum transfer, is shown in Figure 2a for relatively small momentum transfer (large impact parameters). As in previous measurements for purely one-electron transitions, most of the intensity is found in two crescent-shaped areas extending from the target to the projectile, which are separated by a line of minimum intensity at $v_{x,e} = 0$. This indicates a dominant contribution from the quasi-molecular π state, which is even better visible in the transverse plane (see Figure 2b). However, here, for a two-electron transition, the intensity of the two areas differs and a third crescent-shape feature with higher $v_{x,e}$ but much lower probability is seen, which is not present for single electron processes. This third feature indicates contributions from more quasi-molecular states. Figures 2c,d show the calculated electron distribution for the best fit of the phases indicating excellent agreement between the results of the experiment and the calculations. As noted above, remarkably, only two dominant states (1sσ$_g$ and 2pπ$_u$) are required in the 2eHC approach to reproduce the dominant features of the spectrum in the plane perpendicular to the direction of impact (Figure 2b and 2d). The RLTDSE calculations then used $b = 1.4$ a.u. (for $b > R_\delta$), corresponding to a relatively small momentum transfer.

(This impact parameter is the stationary point of the integrand of the Fourier transform for the main promotion channel, 2pπ$_u$, for the scattering angle 1.25 mrad and the internuclear distance of 10 a.u.)

To achieve our goal, the experimental confirmation of zeros in the wave function and corresponding vortices in the electronic current, we therefore have to analyze a different dynamical regime. Close collisions are known to transfer more angular momentum from the nuclear to electronic motion. Such significant changes of the electron angular momentum is decisive for the generation of vortices as has been shown by the RLTDSE method. Therefore, we selected the small fraction of events with large scattering angle from 2.25 mrad to 3.25 mrad from our dataset. The resulting electron distribution in the transverse plane for close collisions (Figure 3a) shows a very sharp peak approximately located at $v_{x,e} = v_{y,e} = 0$ with a width of only 0.04 a.u. The surrounding broader distribution exhibits four local minima. Having confirmed the validity of our 2eHC-RLTDSE method for distant collisions, we next used it to show that these minima have corresponding circulations within the electronic current as depicted in Figure 3b and therefore can be identified as vortices. The 2eHC method required just the two dominant states 2sσ$_g$ and 3dδ$_g$ and the RLTDSE calculations in this case used $b = 0.9$ a.u. ($b < R_\delta$), corresponding to a relatively large momentum transfer. (This impact parameter is the stationary point of the integrand of the Fourier transform for the 3dδ$_g$ channel, for the scattering angle 2.75 mrad and the internuclear distance of 10 a.u.)

Specifically, Figure 3b shows the calculated electronic probability density by the color code and the flux is indicated by arrows that circle around the minima. We conclude that we experimentally found local minima in the momentum distribution of a free electron that are manifestations of vortices. The slight asymmetry of the distribution seen in the experiment is due to the presence of a small contribution from the 2pπ$_u$ channel. In contrast, the theoretical distribution is totally symmetric since it only contains two channels for simplicity.

To explore the correlation between the appearance of vortices and angular momentum transfer from the nuclear motion to the electrons we investigated more closely the quasi-molecular states dominantly contributing to the electron emission pattern shown in Figure 3. As shown by a two-state 2eHC-RLTDSE calculation, the observed structure mainly arises from the coherent superposition of 3dδ$_g$, which has nodes on the diagonals, and 2sσ$_g$, which has a circular node centered at $v_{x,e} = v_{y,e} = 0$. Superimposing these contributions coherently with a phase difference of π/2 results in a distribution with four local minima as we have experimentally observed. However, δ states are populated with only very small probability via one-electron transitions. For the two-electron transition of transfer ionization δ-states are, however, efficiently populated as our 2eHC calculation shows. We find a critical minimum projectile scattering angle (about 2 mrad) necessary for vortices to appear. It corresponds in the stationary phase approximation to the internuclear distance $R_\delta$ at which the $(2p\sigma_u)^2$ state crosses $(1s\sigma_g,3d\sigma_g)$. This is the dominant pathway for the promotion of one electron to a $\delta$-state for collisions with impact parameters smaller than the critical impact parameter $(b \approx R_\delta)$. The united-atom rotational coupling between $3d\sigma_g$ and $3d\delta_g$ gives rise to the vortices seen in Figure 3. The $3d\delta_g$ electron is subsequently promoted to the continuum while the other electron remains bound in the 1s orbital of the projectile. The full set of measurements, made over a wide range of momentum transfers, unambiguously shows this threshold for vortex production.

In conclusion, we have experimentally demonstrated the existence of zeros in the electronic probability density, resulting from vortices in the electronic probability current, for a free electron by detecting the associated minima in the momentum distribution. The vortices result from an angular momentum exchange between the nuclear and electronic motions that arises from a two-electron transition in the breaking of the quasi-molecular bond formed transiently in the process of transfer ionization in slow $He^{2+}$ + He collisions. Moreover, the vortex formation occurs only for close-collisions (here, for scattering angles greater than 2 mrad) corresponding to large momentum transfers. The concordance of theory and experiment helps validate the emerging understanding of the roles played by vortex formation in the collision and those that persist to asymptotic distances. In addition, we have shown that the measured electronic spectra image directly the $\delta$-state that is populated in the corresponding two-electron transitions.

SYO and JHM acknowledge support by the Office of Basic Energy Sciences, US Department of Energy through grant DE-FG02-02ER15283 to the University of Tennessee and SYO acknowledges support from Russia's Federal Program "Scientific and Educational Manpower for Innovative Russia" (Project No. 8526). The experimental work is supported by the Deutsche Forschungsgemeinschaft (SCHM 2382/2-1).

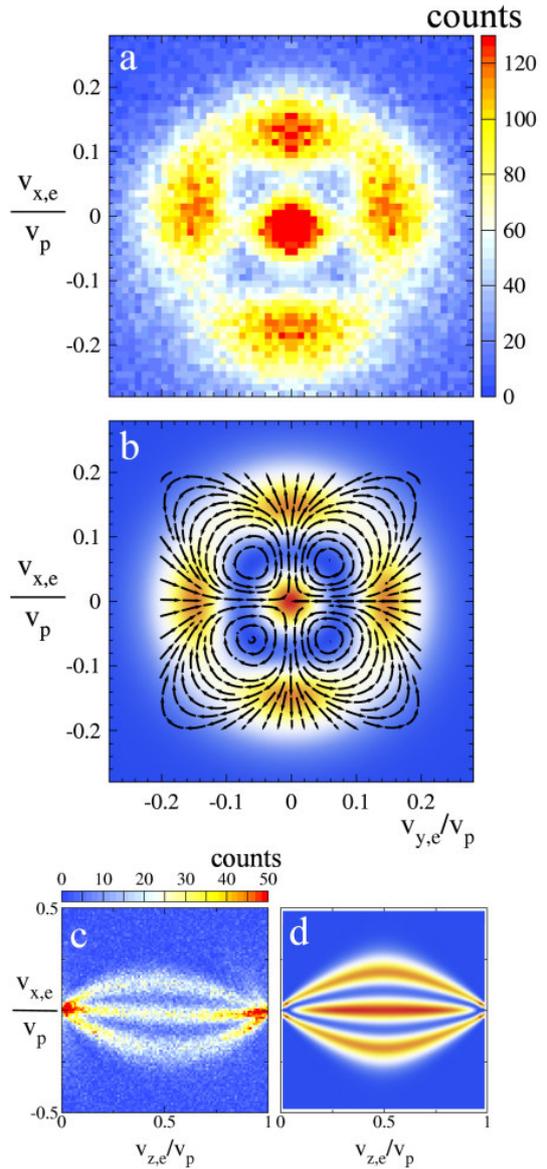

Figure 3. Ejected electron momentum distribution for close collisions. (a) and (c) experimental data for large scattering angles between 2.25 and 3.25 mrad. (b) and (d) calculations for an impact parameter of 0.9 a.u. using only the two states $2s\sigma_g$ and $3d\delta_g$. (a) and (b) show the transverse $(x,y)$ plane with the electronic probability current visualized by arrows shown in the theoretical results (b). The corresponding distributions in the collision plane are shown in (c) and (d).

**References**

[1] R.E. Olson, Phys. Rev. A **27**, 1871 (1983).

[2] R. Dörner, H. Khemliche, M.H. Prior, C.L. Cocke, J.A. Gary, R.E. Olson, V. Mergel, J. Ullrich, and H. Schmidt-Böcking, Phys. Rev. Lett. **77**, 4520 (1996).

[3] M. A. Abdallah, C. L. Cocke, W. Wolff, H. Wolf, S. D. Kravis, M. Stöckli, and E. Kamber, Phys. Rev. Lett **81**, 3627 (1998).

[4] L.Ph.H. Schmidt, M.S. Schöffler, K.E. Stiebing, H. Schmidt-Böcking, R. Dörner, F. Afaneh, and T. Weber, Phys. Rev. A **76**, 012703 (2007).

[5] J. H. Macek and S.Yu. Ovchinnikov, Phys. Rev. Lett. **80**, 2298 (1998).

[6] J. H. Macek, J.B. Sternberg, S.Yu. Ovchinnikov, T.-G. Lee, and D.R. Schultz, Phys. Rev. Lett. **102**, 143201 (2009).

[7] S.Yu. Ovchinnikov, J.H. Macek, L. Ph. H. Schmidt, and D.R. Schultz, Phys. Rev. A **83**, 060701(R) (2011).

[8] L.Ph.H. Schmidt, F. Afaneh, M. Schöffler, J. Titze, O. Jagutzki, Th. Weber, K.E. Stiebing, R. Dörner, and H. Schmidt-Böcking, Phys. Scripta **T110**, 379 (2004).

[9] The impact parameter can only be roughly inferred from the measured scattering angle, so to estimate the uncertainty of experimental impact parameter determination we calculated the impact parameter range contributing to a specific scattering angle by using the CTMC method. We found, for example, that $b = 0.8$ a.u. to 1.5 a.u contribute to a scattering angle of 1.7 mrad.

[10] E.A. Solov'ev, Phys. Rev. A **42**, 1331 (1990).

[11] G.N. Ogurtsov, S.Yu. Ovhinnikov, J.H. Macek, and V.M. Mikoushkin, Phys. Rev. A **84**, 032706 (2011).

[12] D.R. Schultz, M.R. Strayer, and J.C. Wells, Phys. Rev. Lett. **82**, 3976 (1999).

[13] T.-G. Lee, S.Yu. Ovchinnikov, J. Sternberg, V. Chupryna, D.R. Schultz, and J.H. Macek, Phys. Rev. A **76**, 050701 (2007).

[14] S.Yu. Ovchinnikov, G.N. Ogurtsov, J.H. Macek, and Yu.S. Gordeev, Phys. Rep. **389**, 119 (2004).

[15] J. H. Macek, Dynamical Processes in Atomic and Molecular Physics, edited by G. Ogurtsov and D. Dowek (Bentham Science, Oak Park, 2012).